\documentclass[preprint,12pt]{elsarticle}
\usepackage{natbib}
\usepackage{amssymb}
\usepackage{amsmath}
\usepackage{amssymb,amsmath}
\usepackage{epstopdf}
\usepackage{verbatim}
\usepackage{caption}
\usepackage{color}
\usepackage{makecell}
\usepackage{multirow}
\usepackage{graphicx,bm}
\journal{Physics Letters A}

\begin{document}

\begin{frontmatter}


 %%\author{Name\corref{cor1}\fnref{Label2}}
 %%\ead{}

\title{Non-Floquet oscillations of a parametrically driven rigid planar pendulum} %% Article title

\author[]{Rebeka Sarkar}
 \author[]{Krishna Kumar\corref{cor1}}
 \ead{kumar.phy.iitkgp@gmail.com}
 \author[]{Sugata Pratik Khastgir}
 \affiliation[]{organization={Department of Physics, Indian Institute of Technology Kharagpur},
             city={Kharagpur-721302},
 %           postcode={721302},
             state={West Bengal},
             country={India}}
%\author{Rebeka Sarkar} %% Author name
%\author{Krishna Kumar}
%\author{Sugata Pratik Khastgir}
%
%%% Author affiliation
%\affiliation{organization={Department of Physics, Indian Institute of Technology Kharagpur},%Department and Organization
%            city={Kharagpur},
%            postcode={721302}, 
%            state={West Bengal},
%            country={India}}
%% Abstract
\begin{abstract}
The linear and nonlinear motions of a damped rigid planar pendulum, driven by vibrating its pivot sinusoidally, are reexamined. The pendulum is known to exhibit periodic, quasiperiodic, and chaotic motions. Floquet analysis identifies regions of instability and stability within the driving parameter space. A new type of nonlinear oscillation may occur at driving parameters where Floquet analysis predicts a stable stationary state. Such non-Floquet oscillations always have periods longer than twice the period of the vibrating pivot. The possible periods of these oscillations may be four, six, eight, or twelve times the driving period. The power spectrum of the pendulum's angular velocity during these oscillations reveals a novel feature: the two dominant response frequencies sum to the driving frequency.
\end{abstract}


\begin{highlights}
\item Research highlight 1: A parametrically driven rigid pendulum exhibits a novel type of nonlinear oscillations with a period longer than twice the driving period. The new type of oscillations occurs at the driving parameters for which the Floquet stability analysis predicts a stable stationary state.
    
\item Research highlight 2:  The frequencies of the two most dominant peaks in the power spectrum of the angular velocity of the swinging pendulum sum to the drive frequency. This result is analogous to the energy conservation condition in spontaneous parametric down-conversion in quantum optics.  
\end{highlights}

%% Keywords

\end{frontmatter}

%\bibliographystyle{apsrev4-1} % Tell bibtex which bibliography style to use
%\bibliography{xampl} % Tell bibtex which .bib file to use (this one is some example file in TexLive's file tree)


\section{Introduction} 
A planar pendulum is a well-known classical system~\cite{Huygens_book_2005} with a single degree of freedom. It must be driven externally to sustain its oscillations against dissipative forces from air resistance and friction at the pivot. The pendulum exhibits a rich variety of dynamic states under parametric driving~\cite{Stephenson_1908,Stephenson_1909,McLaughlin_1981,Starrett-Tagg_1995,Butikov_1999,Fameli_etal_1999,Bartucelli_2001,Champneys_2009,Das&Kumar_2015,Depetri_etal_2018,Afzali_etal_2021,Lubarda&Lubarda_2021}, where its angular frequency is modulated periodically. The pendulum shows parametric resonance~\cite{Faraday_1831}, which occurs in various fields of science and engineering. For example, they are known to occur in problems of continuum mechanics~\cite{Faraday_1831,Fauve_etal_1992,Edwards-Fauve_1993,Kumar-Tuckerman1994,Kumar_1996,Smorodin_Luecke_2010}, discrete mechanics~\cite{Mechanics_Landau-Lifshitz_1960,Verhulst_2022,Sarkar_etal_2023}, engineering~\cite{Fujii_2006,Krylov_etal_2010,Belykh_etal_2021,Li_etal_2022}, quantum field theory~\cite{Berges&Serreau_2003}, quantum optics~\cite{Leonardo_etal_2007,Svidzinsky_etal_2013}, and condensed matter physics~\cite{Wustman&Shumeiko_2013}, among others. A parametrically driven rigid planar pendulum has two distinct physical stationary states: (i) the pendulum at rest with its bob vertically below the pivot (normal state), and (ii) the pendulum at rest in the inverted state. Floquet analysis~\cite{Floquet_1883} of the linearised equation of motion about its normal state reveals multiple resonance frequencies and tongue-shaped zones of instability in the plane of driving parameters (amplitude $A$ and frequency $\Omega$). If the driving period is $T=2\pi/\Omega$, the pendulum at rest can be excited to perform either subharmonic oscillations of period $2T$ (harmonic oscillations of period $T$) when the driving parameters are chosen from one of the subharmonic (harmonic) instability zones. For driving parameters outside any instability zone, the pendulum will come to rest. This means the normal stationary state is stable against small perturbations, as per the Floquet theory. Similarly, the results of the Floquet analysis of an inverted pendulum identify the range of driving parameters that stabilise it in the inverted state against small perturbations. Although the Floquet theory is strictly valid for linear differential equations with periodic coefficient(s), in practice it is observed that the nonlinear periodic solutions follow the Floquet predictions at least just above the onset of oscillatory motions~\cite{Faraday_1831,Fauve_etal_1992,Edwards-Fauve_1993,Smorodin_Luecke_2010,Sarkar_etal_2023}. Floquet theory predicts oscillations with a period of either $2T$ or $T$. We have examined numerically the linear and nonlinear behaviour of a driven damped pendulum. To our surprise, we discovered that the pendulum may be excited to perform nonlinear periodic oscillations for driving parameters outside any instability zone identified by the Floquet analysis. For these parameters, the Floquet (linear) theory predicts the pendulum should be in a stable, stationary state. The observed non-Floquet-type oscillations are fully nonlinear and exhibit a novel feature: the frequencies of the two largest peaks in the power spectrum sum to the driving frequency. 
They have a period equal to an even multiple of the driving period $T$ but always greater than $2T$. Additionally, an inverted pendulum under parametric driving~\cite{Kapitza_1951} may display subharmonic oscillations of small amplitude.

\section{Equation of motion}
\begin{figure*}[h]
	\centering
	\includegraphics[height=!, width=0.5\textwidth]{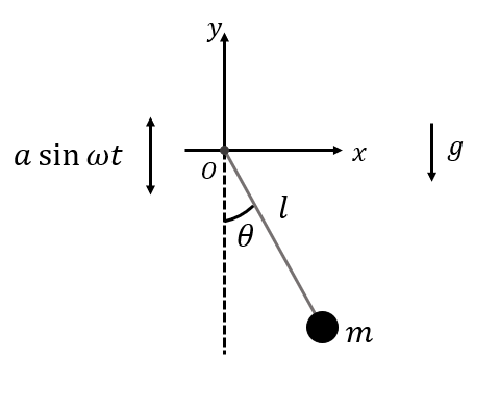}
	\caption{Schematic diagram of a planar pendulum with its pivot vibrated in the vertical direction.}
	\label{schematic}
\end{figure*}
Figure~\ref{schematic} schematically shows a planar pendulum of light rigid rod of length $l$ and a heavy bob of mass $m$, which can oscillate or rotate about a pivot in the vertical ($xy$-) plane under the influence of gravity, $g$. The pivot performs sinusoidal vibration with an amplitude $a$ and angular frequency $\omega$ in the vertical direction about its mean position located at `O'. The mass of the light rod is ignored. The angular displacement of the pendulum at any time $t$  is denoted by the symbol $\theta (t)$. The Lagrangian  $\mathcal{L} (\theta, \dot{\theta}, t)$ of the system~\cite{Mechanics_Landau-Lifshitz_1960}, in the absence of dissipation, then reads as:
\begin{equation}
	\mathcal{L}(\theta, \dot{\theta}, t)=\frac{1}{2}m l^2 \dot{\theta}^2 + m g l \cos{\theta} + mla\omega \dot{\theta} \cos{\omega t}  \sin{\theta} + \frac{df}{dt}, \label{SP}
\end{equation}
%where $f(t, a, \omega)=\frac{1}{2}ma^2\omega^2 \cos^2{\omega t} - m g a \sin{\omega t}$ 
where the function $f(t, a, \omega)$ is independent of dynamical variables ($\theta$ and $\dot{\theta}$) and, therefore, $df/dt$ does not contribute to the equation of motion.
We assume the dissipation force $\bf{F}$ to be velocity-dependent. If $\gamma$ is the coefficient of dissipation, then ${\bf F} = -\gamma {\bf v} = -\frac{\partial}{\partial {\bf v}}\left( \frac{\gamma}{2} {\mathrm{v}^2}\right)=-\frac{\partial {\mathcal{F}}}{\partial {\bf v}}$. The function $\mathcal{F}=\frac{\gamma}{2} {\mathrm{v}^2}=\frac{\gamma}{2} [l^2{\dot{\theta}}^2 + 2 al\omega \dot{\theta} \cos{\omega t} \sin{\theta} + h(t)] $ is known as the Rayleigh dissipation function~\cite{Goldstein}.  The generalised dissipative force is then given by ${\bf Q} (\dot{\theta}) = -\hat{\boldsymbol{\theta}}\frac{\partial \mathcal{F}} {\partial  \dot{\theta}}  =-\gamma (l^2\dot{\theta} + al\omega \cos{\omega t} \sin{\theta})\hat{\boldsymbol{\theta}} \equiv Q (\theta, \dot{\theta}, t) \hat{\boldsymbol{\theta}}$, where the unit vector $\hat{\boldsymbol{\theta}}$ is directed along the azimuthal direction. Here $h(t)$ is function of time only and does not contribute to the equation of motion. The Euler-Lagrange equation, in the presence of a generalised dissipative force ${\bf Q}$, is
\begin{equation}
	\frac{d}{dt}\left(\frac{\partial \mathcal{L}} {\partial \dot{\theta}}\right)-\frac{\partial \mathcal{L}}{\partial \theta} = Q.\label{EL}
\end{equation}
The resulting dimensionless equation of motion takes the following form:	
\begin{equation}
	\frac{d^2\theta}{d\tau^2} + 2\beta \frac{d\theta}{d\tau} + \left(1 - A \sin{\Omega \tau} 
	+\frac{2\beta}{\Omega} A  \cos{\Omega \tau }\right)\sin{\theta} =0,  \label{EqSP}
\end{equation}
where $2\beta = \gamma/(m\omega_0)$ the damping coefficient and $\omega_0 =\sqrt{g/l}$. The dimensionless driving amplitude and angular frequency are $A = a\omega^2/g$ and $\Omega = \omega/\omega_0$, respectively. The dimensionless time is given as $\tau = \omega_0 t$. 
The limit, $2\beta/\Omega \rightarrow 0$ in Eq.~\ref{EqSP}, may be achieved in two ways: (i) $\beta \rightarrow 0$, Eq.~\ref{EqSP} reduces to an undamped driven planar pendulum,  and (ii) $\Omega \rightarrow \infty$,  Eq.~\ref{EqSP} reduces to a damped driven planar pendulum at high frequency.

\begin{figure*}[htb]
	\centering
	\includegraphics[height=!, width=0.95\textwidth]{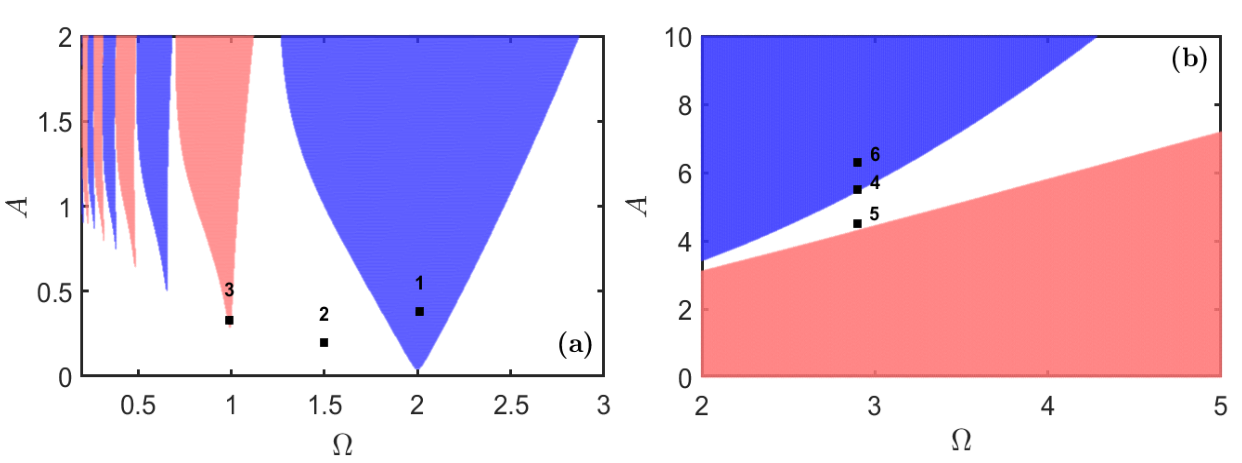}
	\caption{(Colour online) Instability regions in the $\Omega$-$A$ plane for a parametrically  driven rigid planar pendulum with damping coefficient $\beta =0.01$ according to the Floquet analysis.  Instability zones are for stationary states corresponding to a pendulum (a) in its normal state ($\theta^*=\dot{\theta}^*=0$) and (b) an inverted pendulum ($\theta^*=\pi$, $\dot{\theta}^*=0$). Blue (grey) and peach (light grey) coloured regions correspond to subharmonic and harmonic instability. White regions suggest stable stationary states. Driving parameters for points marked: (1) $\Omega=2.01$, $A=0.38$; (2) $\Omega=1.50$, $A=0.20$; (3) $\Omega=0.99$, $A=0.33$; (4) $\Omega=2.90$, $A=5.50$; (5) $\Omega=2.90$, $A=4.50$; (6) $\Omega=2.9$, $A=6.31$.}
	\label{FM}
\end{figure*}

\section{Stability diagram of a driven rigid planar pendulum}
The normal stationary state of the pendulum is defined as: $ \theta^{*}= 0$ and $\dot{\theta}^{*}=0$, while the inverted state is identified as:  $\theta^{*}= \pi$ and $\dot{\theta}^{*}=0$. The former state is stable, while the latter is always unstable in the absence of the driving force. To investigate how the perturbations grow or decay over time if the pendulum is slightly disturbed around its fixed points with a vibrating pivot, we set $\theta = \theta^{*} + \Theta$ and $\dot{\theta}= \dot{\theta}^{*} + \Psi$. Here, $\Theta$ and $\Psi$ are small perturbations around the fixed points. The following two-dimensional linear dynamical system describes the linear stability of the planar pendulum at these fixed points:
\begin{equation}
	\frac{d\mathbf{\Phi}}{d\tau}=M(\tau) \mathbf{\Phi}, \label{ds_SP}
\end{equation}
where ${\mathbf \Phi}$ is a column vector and $M (\tau)$ is a time-periodic $2\times 2$ square matrix defined as:
\begin{equation}	
	{\mathbf \Phi}=\begin{bmatrix}
		\Theta\\ \Psi
	\end{bmatrix},~~~M(\tau)=\begin{bmatrix}
		0 & 1   \\ j\left(1-A\sin{\Omega \tau} + \frac{2\beta }{\Omega}A\cos{\Omega \tau}\right) & -2\beta  
		\quad
	\end{bmatrix}.\label{SPmatrix}
\end{equation}
Equation~\ref{ds_SP} with \ref{SPmatrix} is the modified Mathieu equation~\cite{Mathieu_1868} due to damping. The integer $j$ can have two values: $-1$ for the normal state and $+1$ for the inverted state. Following Floquet theorem~\cite{Floquet_1883}, the solution of Eq.~\ref{ds_SP} can be expressed as a product of an exponential term $e^{\mu \tau}$ and a periodic function $\mathbf{F}(\tau)$. This means, $\mathbf{F}(\tau + T) = \mathbf{F}(\tau)$, where $T=2\pi/\Omega$ is the dimensionless driving period. The solution of Eq.~\ref{ds_SP} may be expressed as $\mathbf{\Phi}(\tau) = e^{\mu \tau} \mathbf{F}(\tau) = e^{\mu \tau} \sum_{-\infty}^{\infty} \mathbf{F}_n e^{in\Omega \tau}$. As time $\tau \rightarrow \tau+T$, the solution $\mathbf{\Phi}(\tau) \rightarrow$ $\mathbf{\Phi}(\tau+T) = e^{\mu (\tau+T)} \mathbf{F}(\tau+T) = e^{\mu (\tau+T)} \mathbf{F}(\tau) = e^{\mu T} \mathbf{\Phi}(\tau)$. The Floquet exponent $\mu$ can be expressed as $\mu=s+i\alpha \Omega$, where $s$ and $\alpha \Omega$ are the growth rate and frequency of the periodic perturbations. Here, $\alpha$  is a real number.  After one period $T$, the solution $\mathbf{\Phi} (\tau)$ is multiplied by the factor $\Lambda = e^{\mu T}$, where $\Lambda$ is known as the Floquet multiplier~\cite{Mechanics_Landau-Lifshitz_1960}. 

The angular displacement ${\mathbf \Phi (\tau)}$ is real only if $\alpha$ is either an odd multiple or an even multiple of $1/2$. So other values of $\alpha$ are non-physical for the pendulum motion. The cases with $\alpha= m + 1/2$ correspond to subharmonic solutions and those with $\alpha=m$ correspond to harmonic solutions for $m=0,1,2,3,\cdots$. Expanding the periodic function $\mathbf{F}(\tau)$ in Fourier series, $\mathbf{F}(\tau) = \sum_{-\infty}^{\infty} \mathbf{F}_n e^{i n\Omega \tau}$,  the angular displacement reads as $\mathbf{\Phi}(\tau) = e^{s\tau} \sum_{-\infty}^{\infty}\mathbf{F}_n e^{i (m+n+1/2)\Omega \tau}$ for subharmonic solutions and $\mathbf{\Phi}(\tau) = e^{s\tau} \sum_{-\infty}^{\infty}\mathbf{F}_n e^{i (m+n)\Omega \tau}$ for harmonic solutions. Redefining the $m+n$ as a new integer $n^{\prime}$, one needs to consider only two values of $\alpha$: $\alpha = 1/2$ for subharmonic cases and $\alpha = 0$ for harmonic cases. If the real part of the Floquet exponent,  $s=\ln{|\Lambda|}/T$, is positive (negative), the oscillatory solution will grow (decay) over time. For $\alpha=0$, the period of the Floquet oscillation is equal to driving period $T$ and the oscillations are synchronous (harmonic) with the driving, while for $\alpha=1/2$ the oscillations are subharmonic with a period $2T$. For $s=0$, the perturbations neither grow nor decay with time. We numerically computed the Floquet multipliers by integrating the dynamical system (Eq.~\ref{ds_SP}) for one period of driving with two different initial conditions: 
$\mathbf{\Phi}_1 (0) = \begin{bmatrix}
	1\\ 0
\end{bmatrix}$ and $\mathbf{\Phi}_2 (0) = \begin{bmatrix}
	0 \\ 1 \end{bmatrix}$. A $2 \times 2$ matrix $\mathbf{B}(T)$ $=$ $[\mathbf{\Phi}_1(T), \mathbf{\Phi}_2(T)]$ is formed using the two solutions at $\tau=T$. Note $\mathbf{B}(0)$ is an identity matrix at $\tau = 0$. The matrix $\mathbf{B}(T)$ is the fundamental matrix for the dynamical system. The eigenvalues of $\mathbf{B}(T)$, which is also called a monodromy matrix, are the Floquet multipliers ($\Lambda_1$ and $\Lambda_2$) of the dynamical system~\cite{Chicone_2001}. The product ($\Lambda_1\Lambda_2$) of two Floquet multipliers~\cite{Mechanics_Landau-Lifshitz_1960} is equal to unity for $\beta=0$ and $e^{-2\beta}$ for $\beta\neq0$. Stability zones in the $\Omega$-$A$ plane is computed using a technique discussed for a parametrically driven double pendulum by Sarkar et al.~\cite{Sarkar_etal_2025}.

Figure~\ref{FM} displays the results of linear stability using Floquet method for a planar pendulum driven by pivot vibration with a small damping coefficient ($\beta = 0.01$). Fig.~\ref{FM} (a) illustrates stability of the driven pendulum around the stationary state $\theta^{*}=0$, $\dot{\theta}^{*}=0$. The pendulum starts swinging for driving parameters chosen from the  tongue-shaped instability zones in the $\Omega$-$A$ plane. The blue (grey) and peach (light grey) coloured regions are zones of subharmonic and harmonic instabilities, respectively. The swing amplitude of the pendulum grows exponentially for any set of driving parameters chosen from the instability zones until the nonlinear terms cannot be ignored. For parameters chosen from white coloured region of the $\Omega$-$A$ plane, swing amplitude decays exponentially and the pendulum comes to rest. The fixed point $\theta^{*}=0$, $\dot{\theta}^{*}=0$ is stable. Fig.~\ref{FM} (b) summarises the results of linear stability analysis for an inverted pendulum. Selecting driving parameters from the white-coloured region can stabilise an unstable inverted planar pendulum. For parameters in the blue (grey) and peach (light grey) regions, an inverted pendulum cannot be stabilised to rest. The amplitude of oscillation will grow either subharmonically or harmonically.

\section{Nonlinear oscillations}
%\subsection[short title]{Floquet oscillations}
The motion of a driven damped planar pendulum, based on linearised equations of motion (Eqs.~\ref{ds_SP} and~\ref{SPmatrix}) will grow or decay exponentially over time except at the boundaries of instability zones, which are curves of neutral stability ($s=0$) where perturbations neither grow nor decay for driving parameters on them. The linear analysis predicts the subharmonic or harmonic instability of the pendulum from its stationary states if the driving parameters are chosen from blue (grey) or peach (light grey) coloured regions. As the values of $\theta$ and $\dot{\theta}$ increase and become relatively large, the pendulum motion is governed by nonlinear equations (Eq.~\ref{EqSP}). To investigate the final nonlinear state of the driven planar pendulum, we introduce two variables $\psi=\dot{\theta}$ and $Z=\Omega \tau$, and convert the dynamical system into the following autonomous form:
\begin{eqnarray}
	\dot{\theta} &=& \psi, \nonumber\\
	\dot{\psi} &=& -2\beta \psi - \left(1 - A \sin{Z} +\frac{2\beta}{\Omega} A \cos{Z}\right)\sin{\theta},\nonumber\\ 
	\dot{Z} &=& \Omega.\label{ds2}
\end{eqnarray}
This dynamical system is integrated using the standard fourth-order Runge-Kutta (RK4) method, with small initial values for $\theta$ and $\psi$ around both fixed points: ($0, 0$) and ($0, \pi$). The initial value of the variable $Z$ is always set as $0$.

\subsection{Floquet oscillations}

\begin{figure*}[!]
	\centering
	\includegraphics[height=!, width=1.0\textwidth]{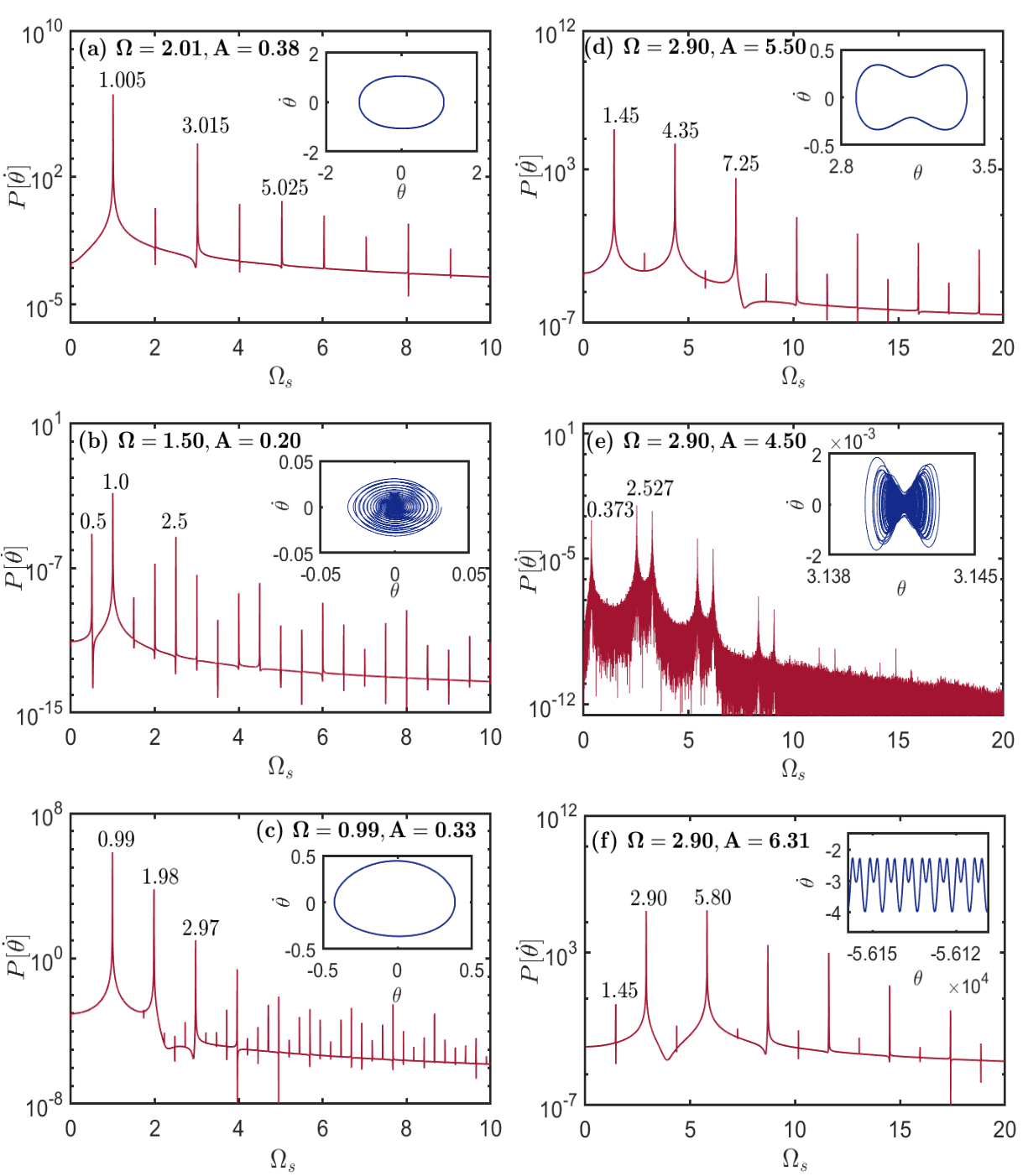}
	\caption{(Colour online) Power spectra of the angular speed $\dot{\theta}$ and phase plots (Insets) of a driven planar pendulum  are shown at damping coefficient $\beta = 0.01$ for different values of driving parameters ($\Omega$, $A$). The left (right) column shows results around the normal (inverted) state of the rigid planar pendulum. Plots in 3(a) \& 3(d) are for subharmonic oscillations and the plots in 3(c) are for harmonic oscillations. Rotational motion is shown in 3(f).  Stable stationary solutions are shown in 3(b) \& 3(e).}
	\label{NM}
\end{figure*}

The results of integration are summarised in Fig.~\ref{NM} for $\beta=0.01$. The left column displays power spectra of the angular speed, $\dot{\theta}$, for different values of driving parameters when it was initially at its normal stationary state ($0, 0$). Insets show the corresponding phase plots. Fig.~\ref{NM}(a) presents the results for $\Omega=2.01$ and $A=0.38$, which lie within a region of subharmonic instability (see point 1, Fig.~\ref{FM}(a)). The largest peak in the power spectrum appears at  $\Omega_s\equiv\Omega_s^{(h1)}=\Omega/2=1.005$. 
The symbol $\Omega_s^{(hn)}$ stand for the frequency of the $n$th dominant peak in the power spectrum of the angular velocity. 
Other significant peaks occur at $\Omega_s = 3\Omega/2, 5\Omega/2, \dots$. Several smaller peaks with lesser heights are also visible at $\Omega_s = \Omega, 2\Omega, 3\Omega, \dots$. The stationary pendulum is excited to perform subharmonic oscillations. Fig.~\ref{NM}(b) shows the pendulum's motion for $\Omega=1.50$ and $A=0.20$, which falls within the white zone (point 2, Fig.~\ref{FM}(a)). The phase trajectories spiral towards the origin ($0, 0$) in the $\theta$-$\dot{\theta}$ plane with a period of $3T$, and the pendulum ultimately comes to rest. The peaks in $P(\dot{\theta})$ are located at $n\Omega/3$, and the sum of the frequencies of the two largest peaks equals $\Omega$. The decaying solution is not Floquet-type. Fig.~\ref{NM}(c) illustrates a case where the driving parameters ($\Omega=0.99$ and $A=0.33$) are just inside the first harmonic instability zone (point 3, Fig.~\ref{FM}(a)). The highest peak in the power spectrum appears precisely at the driving frequency ($\Omega_s^{(h1)}=\Omega$), with other significant peaks at $\Omega_s = 2\Omega, 3\Omega, \dots$. The inset displays the resulting limit cycle corresponding to harmonic oscillations.

The right column of Fig.~\ref{NM} presents results for an inverted pendulum subjected to parametric driving with $\Omega=2.90$ and various values of $A$. When $A=5.50$, which just falls within the blue (grey) region [point 4, Fig.~\ref{FM}(b)], the pendulum oscillates periodically. The limit cycle in the $\theta$-$\dot{\theta}$ plane corresponding to this oscillatory motion is shown in the inset of Fig.~\ref{NM}(d). The clear power spectrum of the angular velocity exhibits peaks at $\Omega_s=(2n-1)\Omega/2$ (for $n=1, 2, 3, \cdots$) with the strongest peak at $\Omega_s=\Omega/2=1.45$. These oscillations are subharmonic, with a period twice that of the driving period. Fig.~\ref{NM}(e) displays results for $A=4.50$, where the parameters fall within the white region [point 5, Fig.~\ref{FM}(b)], indicating small perturbations decay and the pendulum stabilises in the inverted position. A parametrically driven pendulum can be stabilised in this inverted position. The frequencies of the first two  peaks in the power spectrum of the decaying perturbations sum to the driving frequency. Increasing the driving amplitude to $A=6.31$ [point 6, Fig.~\ref{FM}(b)] leads to subharmonic rotational motion. The peaks in the power spectrum, $P[\dot{\theta}]$, are clearly at $\Omega_s=n\Omega/2$ with $n=1, 2, 3, \cdots$, but the dominant frequencies are at $\Omega_s^{(hn)} = n\Omega$. The period of revolution is twice the driving period. 
%A set of six video clips~\cite{Video} attached show the motions described in Fig.~\ref{NM} (a)-(f).

\subsection{Non-Floquet oscillations}

\begin{figure*}[h!]
	\centering
	\includegraphics[height=!, width=0.95 \textwidth]{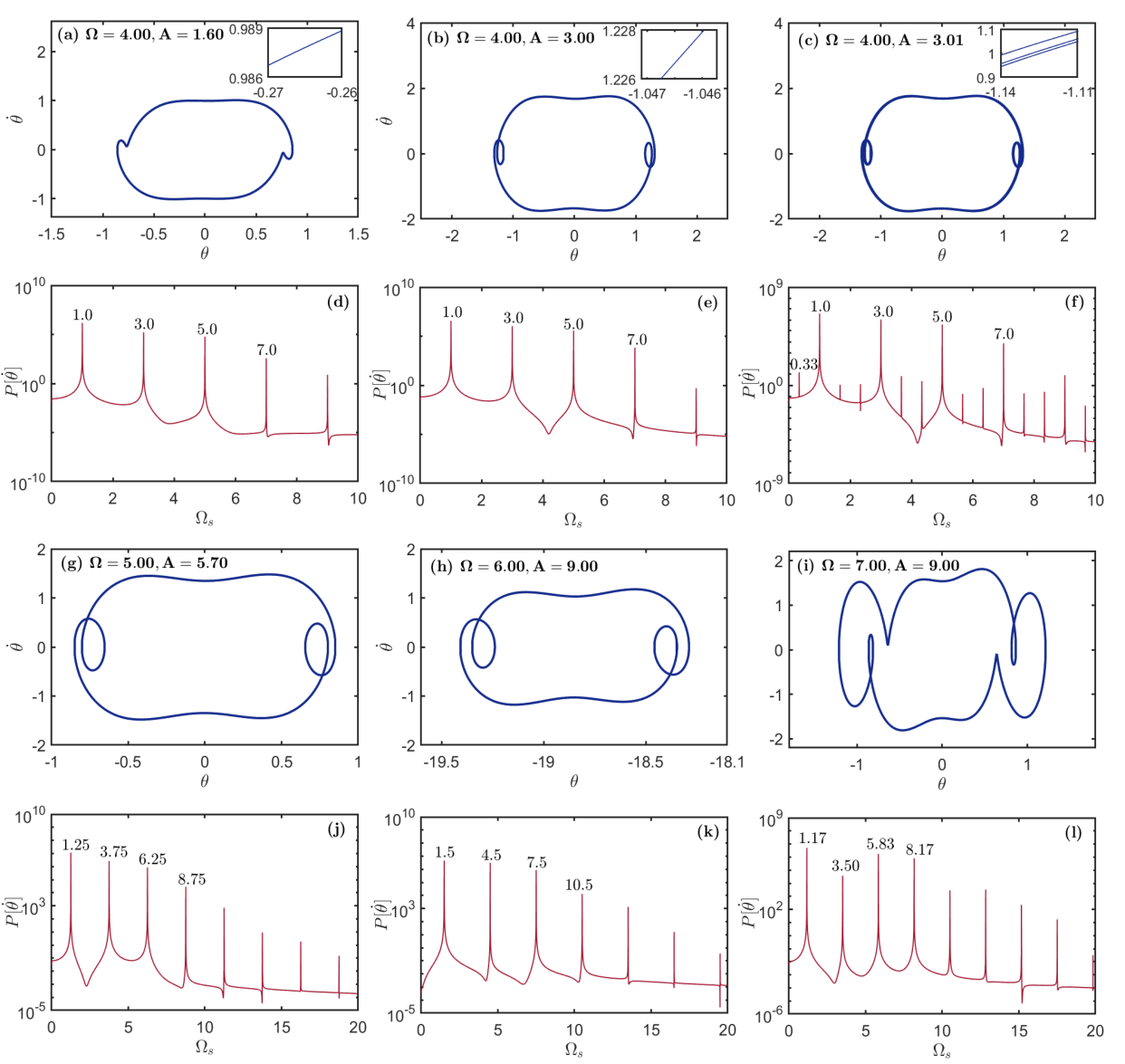}
	\caption{(Colour online) Limit cycles in the $\theta$-$\dot{\theta}$ plane (the first and third rows) for damping coefficient $\beta=0.01$ and corresponding power spectra of the angular speed, $\dot{\theta}$, (the second and fourth rows) for different values of  dimensionless driving frequency, $\Omega$, and amplitude, $A$. (a) $\Omega=4.00$ \& $A = 1.60$, (b) $\Omega=4.00$ \& $A=3.00$, (c) $\Omega=4.00$ \& $A=3.01$, (g) $\Omega=5.00$ \& $A = 5.70$, (h) $\Omega=6.00$ \& $A = 9.00$ and (i) $\Omega=7.00$ \& $A = 9.00$. Insets in (a), (b) and (c) show the number of lines in a small portion of the limit cycles. Plots in (d), (e), (f), (j), (k) and (l) display the power spectra of the angular speed, $\dot{\theta}$, for the limit cycles shown in (a), (b), (c), (g), (h) and (i), respectively.}
	\label{NF1}
\end{figure*}

For values of driving parameters ($\Omega$ and $A$) located outside any of the instability zones (the white-coloured region) of (Fig.~\ref{FM}(a)), the normal stationary state ($0$, $0$) of the driven pendulum is stable against small perturbations. However, if the chosen values of $\Omega$ and $A$ fall into either in the peach (light grey) or blue (grey) coloured regions of the stability diagram for an inverted pendulum (Fig.~\ref{FM}(b)), the stationary state ($\pi$, $0$) for the inverted pendulum becomes unstable. The driving parameter values used for the plots shown in Fig.~\ref{NF1} are from the white region of Fig.~\ref{FM}(a). All these parameters except those for Fig.~\ref{NF1}(h) are also from the peach (light grey) region of Fig.~\ref{FM}(b). Based on Floquet analysis of the linearised equations of a damped planar pendulum under parametric driving, we expect the pendulum to eventually return to its normal state ($0$, $0$) after some time in such cases. This happens especially when perturbations are small.  Driving parameters for Fig.~\ref{NF1}(h) are from the white region of Fig.~\ref{FM}(b). For this case, depending on initial conditions, the pendulum should stay either in normal state or inverted state. 
For larger perturbations, a new type of oscillatory motion appears. Fig.~\ref{NF1} illustrates the nonlinear oscillatory motion for $\beta=0.01$ with various sets of $\Omega$ and $A$. The top row of plots shows the limit cycles for $\Omega=4.00$ and three different $A$ values. The second row presents the corresponding power spectra of the angular velocity $P(\dot{\theta})$. For $A=1.60$, the pendulum swings periodically, forming the limit cycle shown in Fig.~\ref{NF1}(a). The power spectrum  $P(\dot{\theta})$ (Fig.~\ref{NF1}(d)) exhibits peaks at $\Omega_s \equiv \Omega_s^{(hn)}=(2n-1)\Omega/4$, where $ n=1,2,3,4, \ldots$. The  first peak, which is the largest of all peaks, is located at one-quarter of the driving frequency. The pendulum oscillates with a period four times the driving period ($T$). This nonlinear oscillation is not predicted by Floquet analysis. As $A$ increases slightly, the size and shape of the limit cycle change. At $A=3.00$, the limit cycle develops a loop-like path (Fig.~\ref{NF1}(b)) near the points of maximum deflection. The power spectrum again peaks at $\Omega_s=(2n-1)\Omega/4$, with the period remaining $4T$ for $A$ between $1.60$ and $3.00$, assuming fixed $\beta$ and $\Omega$. In both cases, the limit cycles are single lines (see insets, Fig.~\ref{NF1}(a) and(b)). The frequencies of the two largest peaks sum to the driving frequency ($\Omega_s^{(h1)} + \Omega_s^{(h2)} = \Omega$). This motion is extremely sensitive to initial conditions. Slight increase in $A$ keep the limit cycle shape similar, but the phase path shows multiple lines. For $ A=3.01$, Fig.~\ref{NF1}(c) shows a limit cycle with three closely spaced lines, indicating spontaneous period tripling from $4T$ to $12T$. The power spectrum (Fig.~\ref{NF1}(f)) with peaks at frequencies $\Omega_s = (2n-1)\Omega/12$ ($n=1, 2, 3, \cdots$) confirms this. The first peak is located at $\Omega_s=\Omega/12$ and it is not the dominant one. The two largest peaks located at $\Omega_s^{(h1)}=1$ and $\Omega_s^{(h2)}=3$, and these two frequencies still add to the driving frequency ($\Omega=4$).

The third row of Fig.~\ref{NF1} displays the limit cycles for three different sets of $\Omega$ and $A$, while the fourth row shows the corresponding power spectra of the angular velocity. Fig.~\ref{NF1}(g) depicts a limit cycle for $\Omega=5.00$ and $A=5.70$ with an oscillation period of $4T$, confirmed by its power spectrum [Fig.~\ref{NF1}(j)]. The limit cycle for $\Omega=6.00$ and $A=9.00$ [Fig.~\ref{NF1}(h)] is a special case; the driving parameters fall in the white regions of Fig.~\ref {FM}(a) \& (b). This indicates that the normal and inverted states are stable for appropriately chosen small deviations from the equilibrium points. For initial condition $\theta=\pi/10, \dot{\theta}=0$, which is close to the equilibrium point ($0, 0$), the pendulum settles into its normal state after some time. Further, with initial condition $\theta=3.13, \dot{\theta}=0$, which is close the equilibrium point ($\pi, 0$), the pendulum can be stabilised in the inverted state. However, for initial conditions sufficiently away from the two stationary states (e.g., $\theta=\pi/4$ and $\dot{\theta}=0$), the pendulum does not settle in either state; instead, it first rotates about the pivot and then oscillates around the normal equilibrium point ($-6\pi, 0$). The corresponding limit cycle [Fig.~\ref{NF1}(h)] and the power spectrum [Fig.~\ref{NF1}(k)] exhibit features similar to those discussed above. In all these cases, the dominant peak in the power spectrum of the angular velocity occurs at $\Omega_s=\Omega/4$, indicating an oscillation period of $ 4T$ when it is the first peak. If the first peak appears at a lower frequency, as in Fig.~\ref{NF1}(c), the period is longer. The limit cycle for $\Omega=7.00$ and $A=9.00$ is shown in Fig.~\ref{NF1}(i). The most prominent peak of its power spectrum appears at $\Omega_s=\Omega/6=1.1666\approx1.17$, signifying oscillations with a period of $ 6T$ [Fig.~\ref{NF1}(l)]. Other peaks are located at $\Omega_s = (2n-1)\Omega/6$. All the limit cycles shown in Fig.~\ref{NF1} demonstrate inversion symmetry about the normal fixed point.

\begin{figure*}[h]
	\centering
	\includegraphics[height=!, width=0.88 \textwidth]{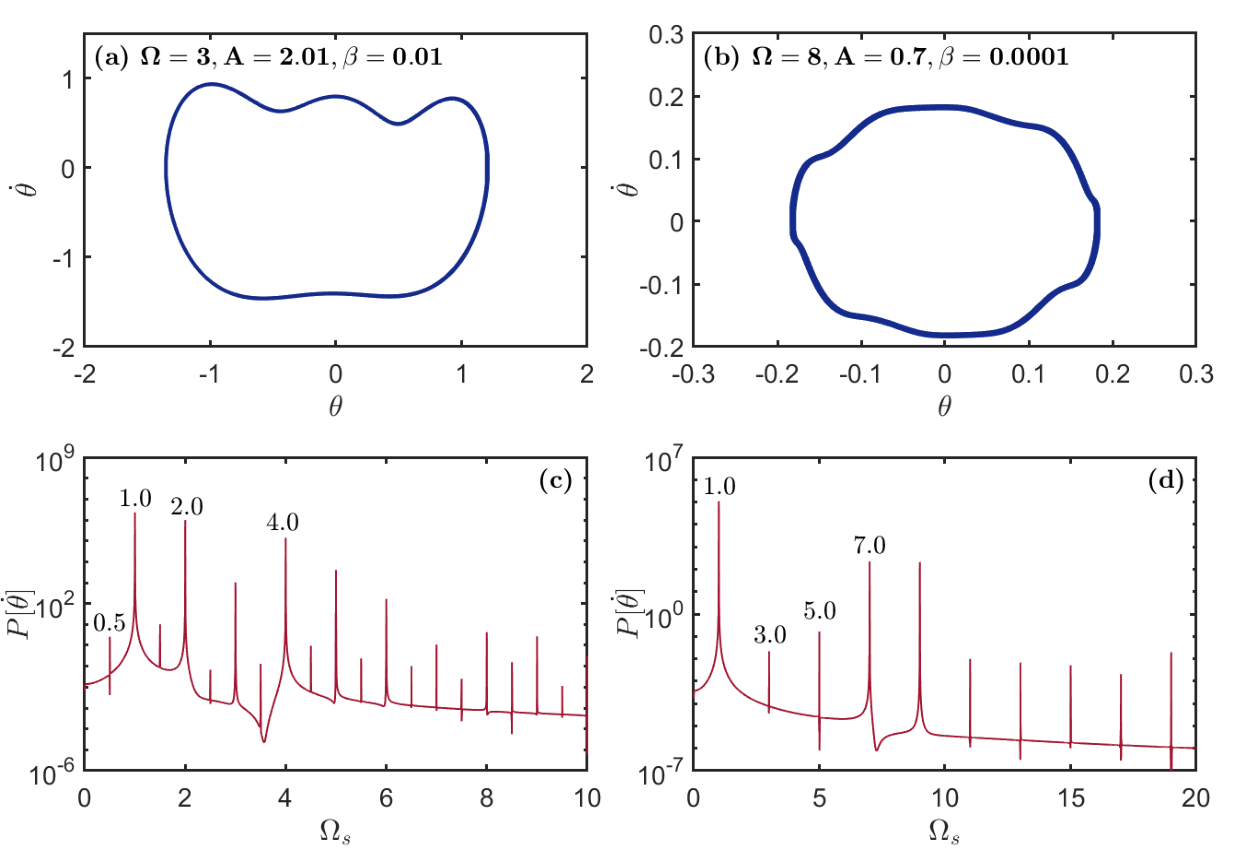}
	\caption{(Colour online) Limit cycles in the $\theta$-$\dot{\theta}$ plane (top row) and power spectra of the angular velocity $P(\dot{\theta})$ (bottom row). Parameters are: $\Omega =3.0$, $A=2.01$ and $\beta=0.01$ for (a) \& (c); $\Omega =8.0$, $A=0.70$ and $\beta=10^{-4}$ in (b) \& (d).}
	\label{NF2}
\end{figure*}

The shape of the limit cycles for non-Floquet oscillations is significantly different for $\Omega=3$ and $8$. Fig.~\ref{NF2}(a) illustrates an oscillatory motion  for $\Omega=3.0$, $A=2.01$ and $\beta=0.01$. There are several peaks in the corresponding power spectrum $P(\dot{\theta})$ [Fig.~\ref{NF2}(c)]. The peaks appear at frequencies, which are multiple of one-sixth times the driving frequency [$\Omega_s=n\Omega/6$ ($n=1, 2, 3, \cdots$)]. The most dominant peak is at $\Omega_s^{(h1)}=2\Omega/6=1$. Therefore, the period of oscillation is six times the driving period. The limit cycle consists of two closely spaced lines. Unlike other limit cycles, it does not possess inversion symmetry about the normal equilibrium point. However, the sum of two frequencies corresponding to the first two largest peaks, located at $\Omega_s=1$ and $2$, is again equal to the driving frequency, $\Omega=3$.  Fig.~\ref{NF2}(b) shows a limit cycle for very small dissipation ($\beta=10^{-4}$) at $\Omega=8.0$ and $A=0.70$. The power spectrum, $P(\dot{{\theta}})$, shown in Fig.~\ref{NF2}(d) is very clean.
The peaks are at $\Omega_s=(2n-1)\Omega/8$ ($n=1, 2, 3, \cdots$).
The oscillatory motion has a period of $8T$, as the first and the largest peak is at $\Omega_s^{(h1)} = \Omega/8$.  The sum of frequencies corresponding to the two largest peaks located at $\Omega_s^{(h1)}=1.0$ and $\Omega_s^{(h2)}=7.0$ again equals $\Omega$.

In all the cases of non-Floquet oscillations (Figs.~\ref{NF1} \& \ref{NF2}), the sum of the frequencies corresponding to the two largest peaks adds to the driving frequency. This novel feature is observed for the first time in a parametrically driven planar pendulum or any classical system with one degree of freedom. All limit cycles correspond to subharmonic oscillations. The driven pendulum oscillates, depending on driving parameters, with a period always greater than twice the driving period. 

\subsection{Model functions}
\begin{figure*}[h!]
	\centering
	\includegraphics[height=!, width=0.90\textwidth]{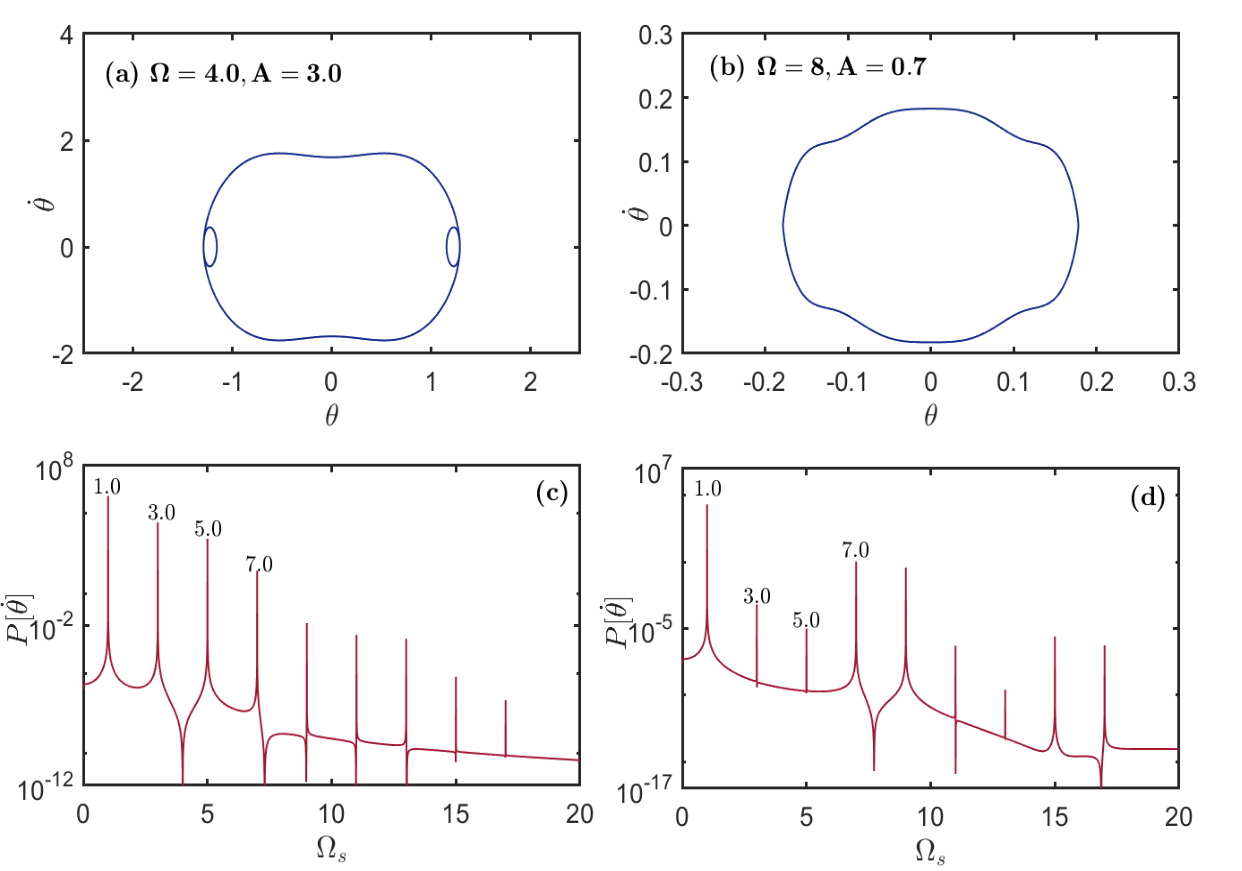}
	\caption{(Colour online) Limit cycles in the $\theta$-$\dot{\theta}$ plane (top row) and power spectra of the angular velocity (bottom row) for model functions with $\beta=0$. Other parameters are: $\Omega = 4.0$ and $A=3.0$ [(a) \& (c)]; $\Omega = 8.0$ and  $A=0.70$ [(b) \& (d)].}
	\label{model}
\end{figure*}

To model the solution $\theta$, the function $\sin{\theta}$ can be approximated by writing
\begin{equation}
	\sin{\theta}\approx \theta -\theta^3/3! + \theta^5/5!
\end{equation}	
and then expanding the generalised coordinate $\theta (\tau)$ up to a suitable higher harmonics. Figure~\ref{model} shows limit cycles for (a) $\Omega = 4.0$ \& $A=3.0$ and (b) $\Omega = 8.0$ \& $A=0.70$. To compute the limit cycle from the full equation (see, Fig.~\ref{NF1}(b)), we expand $\theta (\tau)$ up to seventeenth harmonic as:
\begin{eqnarray}
	\theta(\tau) &=& 1.42673 \cos{\tau} - 0.21078 \cos{3\tau} - 0.0666734 \cos{5\tau} + 0.0065067 \cos{7\tau} \nonumber\\&-& 0.0001518 \cos{9\tau} 
	- 0.000065 \cos{11\tau} + 0.0000482 \cos{13\tau}\nonumber\\
	&-& 0.0000032 \cos{15\tau} - 0.0000006 \cos{17\tau}.\label{model1}	
\end{eqnarray}	
This expression satisfies the equation of motion for $\Omega=4.0$ and $A=3.0$ in the absence of damping ($\beta=0$) within an error of the order of $10^{{-4}}$. The phase plot using the model function [Fig.~\ref{model}(a)] closely matches the numerical solution [Fig.~\ref{NF1}(b)]. The power spectra obtained numerically [Fig.~\ref{NF1}(e)] and from the model function [Fig.~\ref{model}(c)] also have similar features.	To model another solution shown in Fig.~\ref{NF2} (b) for $\Omega=8.0$ and $A=0.70$, the generalised coordinate $\theta$ is expanded as:
\begin{eqnarray}
	\theta(\tau) &=& 0.18044 \cos{\tau} - 0.000032 \cos{3\tau} + 0.000004 \cos{5\tau} -
	0. \ 0.00131\cos{7\tau}\nonumber \\
	 &-& 0.000786 \cos{9\tau} 
	+ 0.0000009 \cos{11\tau} + 0.00000002 \cos{13\tau}\nonumber \\
	 &+& 0.00000204 \cos{15\tau} + 0.000000946 \cos{17\tau}.\label{model2}
\end{eqnarray} 
Fig.~\ref{model}(b) shows the limit cycle based on the above model. The error in this case is of the order of $10^{-6}$. The size and shape of both limit cycles match quite well, but the orientation of the two limit cycles slightly differs [see Fig.~\ref{NF2}(b) \& \ref{model} (b)]. The power spectra from two methods [Fig.~\ref{NF2}(d) \& \ref{model} (d)] are in good agreement. 
We have converged on these forms of expansion (Eqs.~\ref{model1} and~\ref{model2}) with trial model solutions (based on the power spectra of angular velocity, see Fig.~\ref{NF1}(e) and Fig.~\ref{NF2}(d)) in the software Mathematica{\textsuperscript{\textregistered}} by iteratively modifying the amplitudes and minimising the error. Once one chooses first few mode frequencies from the power spectrum for a given driving frequency, other modes  are generated due to the problem's nonlinearity.

\section{Conclusions}
A parametrically driven rigid planar pendulum also exhibits non-Floquet-type oscillations alongside known Floquet-type oscillations. Several systems exhibit Floquet-type nonlinear oscillations with periods $T$ or $2T$, where $T$ is the driving period. The non-Floquet oscillations are essentially nonlinear and cannot be predicted by a linear stability theory. They occur at driving parameters outside instability zones, for which Floquet analysis suggests the pendulum is in a stable, normal state. Non-Floquet oscillations are subharmonic, with a time period always exceeding $2T$. We have presented explicit non-Floquet solutions with possible periods of $4T$, $6T$, $8T$, and $12T$ for different sets of driving parameters. The power spectrum of the angular speed, computed numerically,  exhibits a novel feature: the response frequencies associated with the two largest modes consistently sum to the driving frequency. This feature is analogous to the energy conservation condition observed in spontaneous parametric down-conversion (SPDC) in quantum optics.
\\ 

\noindent{\bf{Acknowledgment}}

Rebeka Sarkar received DST-INSPIRE Fellowship from Govt. of India to carry out this work.
\bigskip

%\noindent{\bf{References}}

\bibliographystyle{elsarticle-num}


\end{document}